\documentclass[pdflatex,sn-mathphys-num]{sn-jnl}

\usepackage{graphicx}%
\usepackage{multirow}%
\usepackage{amsmath,amssymb,amsfonts}%
\usepackage{amsthm}%
\usepackage{mathrsfs}%
\usepackage[title]{appendix}%
\usepackage{xcolor}%
\usepackage{textcomp}%
\usepackage{manyfoot}%
\usepackage{booktabs}%
\usepackage{algorithm}%
\usepackage{algorithmicx}%
\usepackage{algpseudocode}%
\usepackage{listings}%
\usepackage{float}%
\usepackage{placeins}%
\usepackage{xurl}%
\graphicspath{{figures/}}

\theoremstyle{thmstyleone}%
\newtheorem{theorem}{Theorem}
\newtheorem{proposition}[theorem]{Proposition}%

\theoremstyle{thmstyletwo}%
\newtheorem{example}{Example}%
\newtheorem{remark}{Remark}%

\theoremstyle{thmstylethree}%
\newtheorem{definition}{Definition}%

\begin{document}

\title[FuXi-ONS]{Data-driven ensemble prediction of the global ocean}


\author[1,2,4]{\fnm{Qiusheng} \sur{Huang}}

\author[1,5]{\fnm{Xiaohui} \sur{Zhong}}

\author[1,3]{\fnm{Anboyu} \sur{Guo}}

\author[1,3]{\fnm{Ziyi} \sur{Peng}}

\author[1,4]{\fnm{Lei} \sur{Chen}}

\author*[1,2,4,5]{\fnm{Hao} \sur{Li}}\email{lihao\_lh@fudan.edu.cn}

\affil[1]{\orgdiv{Artificial Intelligence Innovation and Incubation Institute}, \orgname{Fudan University}, \orgaddress{\city{Shanghai}, \country{China}}}

\affil[2]{\orgname{Shanghai Innovation Institute}, \orgaddress{\city{Shanghai}, \country{China}}}

\affil[3]{\orgname{Department of Atmospheric and Oceanic Sciences}, \orgname{Fudan University}, \orgaddress{\city{Shanghai}, \country{China}}}

\affil[4]{\orgname{Shanghai Academy of Artificial Intelligence for Science}, \orgaddress{\city{Shanghai}, \country{China}}}

\affil[5]{\orgname{FuXi Intelligent Computing Technology Co., Ltd.}, \orgaddress{\city{Shanghai}, \country{China}}}

\abstract{Data-driven models have advanced deterministic ocean forecasting, but extending machine learning to probabilistic global ocean prediction remains an open challenge. Here we introduce FuXi-ONS, the first machine-learning ensemble forecasting system for the global ocean, providing 5-day forecasts on a global $1^{\circ}$ grid up to 365 days for sea-surface temperature, sea-surface height, subsurface temperature, salinity and ocean currents. Rather than relying on repeated integration of computationally expensive numerical models, FuXi-ONS learns physically structured perturbations and incorporates an atmospheric encoding module to stabilize long-range forecasts. Evaluated against GLORYS12 reanalysis, FuXi-ONS improves both ensemble-mean skill and probabilistic forecast quality relative to deterministic and noise-perturbed baselines, and shows competitive performance against established seasonal forecast references for SST and Ni\~no3.4 variability, while running orders of magnitude faster than conventional ensemble systems. These results provide a strong example of machine learning advancing a core problem in ocean science, and establish a practical path toward efficient probabilistic ocean forecasting and climate risk assessment.}




\maketitle

\section{Introduction}\label{sec1}


The ocean stores more than 90\% of the excess heat trapped by
anthropogenic greenhouse gases \cite{vonSchuckmann2023} and mediates
the planetary redistribution of energy, moisture, and momentum across
scales spanning millimetres to megametres.  Forecasting its evolution
underpins an ever-widening portfolio of societal applications: from
search-and-rescue path planning and oil-spill response on timescales of
hours to days \cite{Barker2020,Melsom2012}, through subseasonal
prediction of marine heatwaves and fishery productivity, to seasonal
outlooks of the El~Ni\~{n}o--Southern Oscillation (ENSO) and Indian
Ocean Dipole that steer agricultural and water-resource decisions
worldwide \cite{Schiller2020}.  Yet the ocean's ``internal weather''---
the turbulent mesoscale eddy field whose characteristic length scale
(the Rossby deformation radius) is an order of magnitude smaller than
its atmospheric counterpart---renders the future ocean state acutely
sensitive to minute perturbations in the present \cite{Lorenz1963,
Lorenz1996}.  A single deterministic forecast, however skilful at short
range, inevitably diverges from reality as nonlinear instabilities
amplify initial-condition errors beyond the predictability horizon
\cite{Thoppil2021}.  What users therefore need is not a single
trajectory but a \emph{probability distribution} over possible futures:
a quantification of the ``error of the day'' that tracks the
flow-dependent growth of uncertainty in real time
\cite{Houtekamer2016,Hoteit2025}.


Numerical weather prediction (NWP) recognised this imperative decades
ago.  Operational atmospheric ensemble prediction systems (EPSs) now
routinely run 50 or more perturbed members, and the resulting
probabilistic products constitute the backbone of hazard early-warning
chains globally \cite{Du2018}.  Ocean forecasting, by contrast, has
been slower to embrace the ensemble paradigm.  The principal obstacle is
computational: resolving the ocean's dominant source of intrinsic
variability---mesoscale eddies---demands horizontal grid spacings of
$\sim$1/12$^{\circ}$ ($\sim$9\,km) or finer, roughly ten times the
resolution at which atmospheric models already capture synoptic-scale
cyclones \cite{Hoteit2025}.  This resolution requirement forces
operational centres into a stark trade-off between the number of
ensemble members and the fidelity of each member.  The European Centre
for Medium-Range Weather Forecasts (ECMWF) runs its OCEAN5/ORAS6
reanalysis--analysis system at 0.25$^{\circ}$ with only 5--11 members
\cite{Zuo2019}; the UK Met Office FOAM system achieves eddy-resolving
1/12$^{\circ}$ but at considerable expense per additional member
\cite{Lea2022}; the US Navy's Earth System Prediction Capability
(ESPC) pushes 16 members to 1/12$^{\circ}$ for subseasonal horizons,
representing perhaps the most computationally aggressive configuration
in current operations \cite{Barton2021}; and Australia's OceanMAPS
augments 48 dynamical members with 144 static (climatological) members
to alleviate sampling error in its EnKF-C assimilation
\cite{Brassington2023}.  Across all these systems, the ensemble sizes
remain one to two orders of magnitude below what would be needed to
adequately sample the tails of high-dimensional ocean-state
distributions.

Equally important is the sophistication of the perturbation strategy.
Modern systems draw on an arsenal of techniques inherited from NWP:
singular vectors \cite{Buizza1995}, breeding vectors \cite{Toth1997},
ensembles of data assimilations (EDA) that perturb observations and
forcing simultaneously \cite{Zuo2019}, stochastic perturbation of
parameterised tendencies (SPPT; \cite{buizza1999stochastic}), stochastic
kinetic-energy backscatter (SKEB; \cite{Shutts2005}), and stochastic
parameter perturbation (SPP) applied to vertical-mixing coefficients and
drag laws \cite{Storto2021}.  Despite their ingenuity, all these
strategies share a common bottleneck: every additional ensemble member
requires a full forward integration of a three-dimensional,
multi-million-grid-cell OGCM coupled to a sea-ice model and driven by a
perturbed atmospheric boundary layer---a cost that scales linearly with
member count and remains, for global eddy-resolving configurations,
prohibitive at large ensemble sizes.

A parallel revolution, meanwhile, has been reshaping geophysical prediction.
Machine-learning (ML) models trained on large reanalysis archives have recently
achieved substantial progress in deterministic forecasting, often delivering
competitive skill at a fraction of the computational cost of conventional
numerical systems \cite{Lam2023_GraphCast,Bi2023_Pangu,Pathak2022_FourCastNet,
Chen2023_FuXi,chen2023fengwu,huang2025fuxi,zhu2026aviasafe}. In oceanography, data-driven
models have likewise shown encouraging performance in forecasting sea-surface
temperature, sea-surface height, subsurface thermal structure, and ocean
circulation across a range of spatial scales
\cite{wang2024xihe,yang2024langya,cui2025forecasting,huang2025fuxiocean}. Together, these advances
suggest that AI has become a viable alternative to conventional numerical
prediction not only in the atmosphere but also in the ocean.

The next step is probabilistic forecasting. In the atmospheric community,
recent studies have shown that AI can be extended from deterministic prediction
to ensemble forecasting, producing calibrated probabilistic forecasts that are
competitive with leading operational systems such as ECMWF ENS
\cite{Price2024_GenCast,zhong2024fuxiens,chen2024machine}. These developments
suggest that a similar transformation may also be possible for ocean
forecasting, where ensemble prediction is equally essential for uncertainty
quantification and risk-aware decision-making.
Yet transferring this success from the atmosphere to the ocean is not
straightforward. Oceanic uncertainty evolves more slowly, is strongly shaped by
subsurface memory, and varies substantially across variables, depths, and
regions. In addition, long-range ocean prediction is more tightly constrained by
the quality of surface forcing, making it difficult to directly extend existing
AI ensemble strategies developed for weather forecasting. A practical
data-driven ocean ensemble system must therefore generate physically meaningful
forecast diversity while remaining stable, coherent, and computationally
efficient over seasonal to annual lead times.

Here we address this gap with \emph{FuXi-ONS}, a data-driven ensemble forecasting
system for the global ocean. FuXi-ONS produces 24-member probabilistic forecasts
of sea-surface temperature, sea-surface height, subsurface temperature,
salinity, and ocean currents on a global $1^{\circ} \times 1^{\circ}$ grid at
5-day intervals out to 365 days. Rather than repeatedly integrating a full
numerical ocean model for each ensemble member, our approach learns both the
evolution of ocean states and the associated uncertainty directly from data.
This enables fast ensemble generation while preserving coherent spatial
variability and physically plausible forecast spread. At a high level, the
system is designed around three key considerations: representing
flow-dependent ocean uncertainty in a data-driven manner, maintaining spatially
structured ensemble diversity across variables and depths, and reducing
sensitivity to long-range atmospheric forcing errors that can otherwise degrade
ocean forecasts at extended lead times.

Our work makes three main contributions. First, to the best of our knowledge,
we present the first AI-based ensemble prediction system for the global ocean,
extending data-driven ocean forecasting from deterministic prediction to
probabilistic prediction. Second, we show that a data-driven ensemble framework
can generate calibrated and physically meaningful forecast spread for
multivariate three-dimensional ocean states at subseasonal to annual lead
times. Third, we demonstrate that such probabilistic forecasts can be produced
with substantially greater computational efficiency than conventional numerical
ensemble systems, highlighting a practical path toward operational
AI-driven ocean ensemble prediction.

\section{Results}\label{sec2}
We evaluate FuXi-ONS on an independent test period from 2021 to 2023, using daily initializations throughout the full record. Unless otherwise stated, all global evaluations in this section are computed over this 2021--2023 test set. For FuXi-ONS, inference is conducted with 24 ensemble members generated on 8 NVIDIA A100 GPUs. The North American MultiModel Ensemble (NMME) \cite{kirtman2014north} reference provides 10 ensemble members in the publicly available forecasts used in this study. For ENSO-oriented seasonal comparison, we additionally consider the forecast products from the International Research Institute for Climate and Society (IRI) \cite{ehsan2024real}, including IRI-D, which uses 18 dynamical models, and IRI-ALL, which further incorporates 8 statistical models. Because the publicly accessible NMME outputs available to us only extend through 2021, all comparisons involving NMME are restricted to forecasts initialized in 2021, using the same verification period and matched evaluation protocol for fairness, following the common practice of comparing on identical initialization sets in subseasonal and ensemble forecast assessment \cite{zhong2024fuxiens,chen2023fuxis2s,Price2024_GenCast}.

To isolate the contribution of the learned ensemble generation mechanism, we compare FuXi-ONS with three reference baselines. FuXi-Aim denotes the deterministic counterpart obtained by removing the noise-generation module and retaining only the direct forecasting backbone. FuXi-Aim-Perlin uses the same deterministic forecast trajectory as FuXi-Aim, but forms an ensemble by adding prescribed Perlin spatial perturbations to the predicted states at inference time. Persistence is constructed by holding the initial ocean state fixed throughout the forecast range and using it as the prediction at all lead times. This comparison setup allows us to distinguish gains brought by learned, state-dependent ensemble generation from those obtained by deterministic forecasting alone, by post hoc stochastic perturbation, or by a no-evolution reference forecast \cite{zhong2024fuxiens,chen2023fuxis2s,Price2024_GenCast}.

We report deterministic metrics for the ensemble mean, including the root mean square error (RMSE) and anomaly correlation coefficient (ACC), together with probabilistic metrics computed from all ensemble members, including the continuous ranked probability score (CRPS), the spread-skill ratio (SSR), and the Ni\~no3.4 index for ENSO-oriented seasonal assessment \cite{barnston1997documentation,l2013linear,fortin2014should}. Unless explicitly noted in the corresponding subsection, the results shown below are averaged over all available forecast cases under the above setting.
\subsection{Overall ocean forecast skill}

\begin{figure}[ht]
\centering
\includegraphics[width=\linewidth]{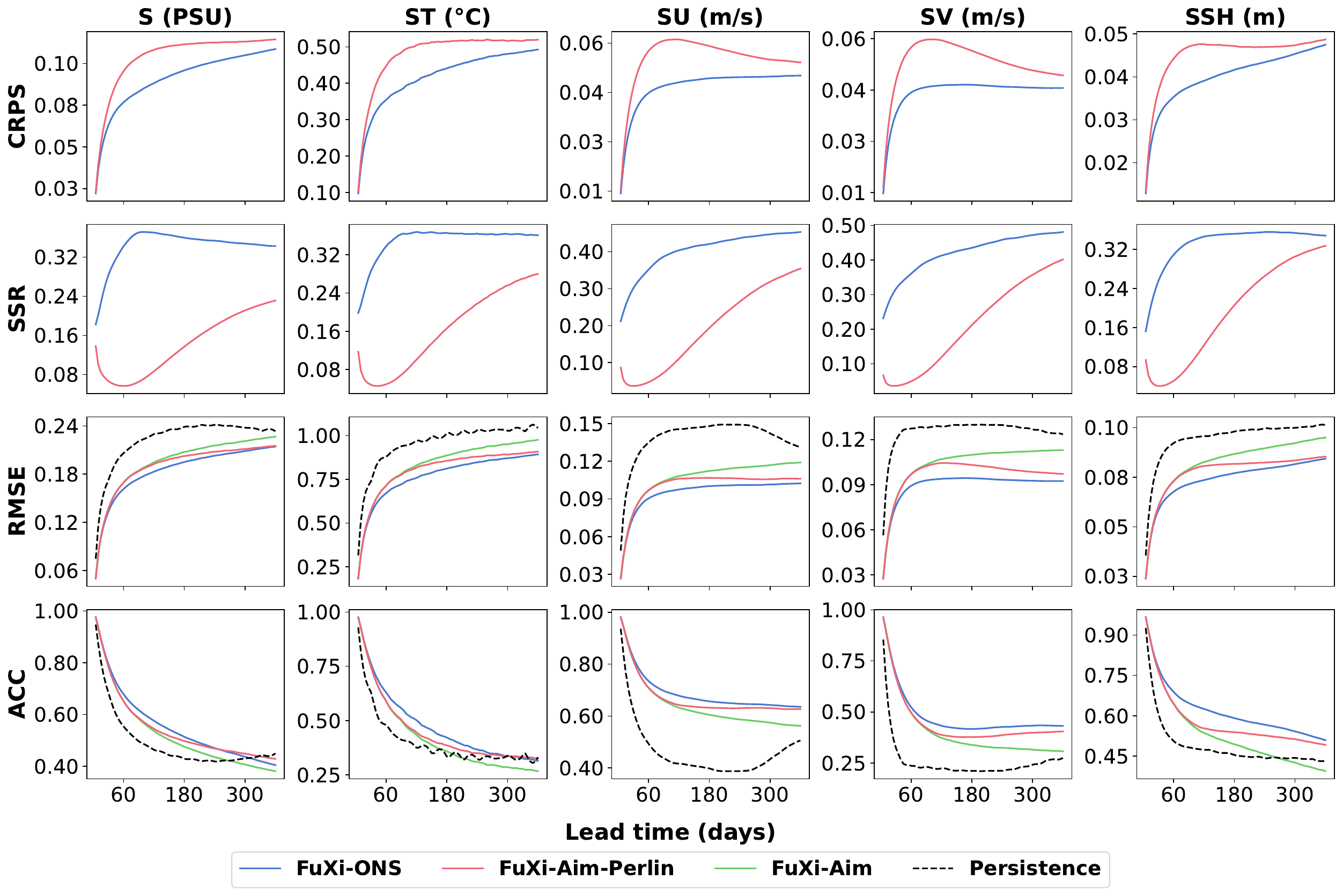}
\caption{\textbf{Depth-averaged forecast skill Comparison of FuXi-ONS with baselines over all lead times.} Forecast performance over the 2021--2023 test period for salinity (S), temperature (ST), zonal current (SU), meridional current (SV), and sea surface height (SSH) as a function of lead time up to 360 days. Rows show the continuous ranked probability score (CRPS), spread-skill ratio (SSR), root mean square error (RMSE), and anomaly correlation coefficient (ACC). CRPS and SSR are reported only for the two ensemble methods, FuXi-ONS and FuXi-Aim-Perlin. RMSE and ACC further include the deterministic baseline FuXi-Aim and persistence. Lower values indicate better performance for CRPS and RMSE, whereas higher values indicate better performance for SSR and ACC. FuXi-ONS consistently provides the best overall probabilistic performance and the strongest deterministic performance of the ensemble mean across most variables and lead times.}
\label{fig_depth_averaged}
\end{figure}
\FloatBarrier
We first evaluate the overall forecast quality of FuXi-ONS using depth-averaged metrics across all predicted ocean variables and lead times (Fig.~\ref{fig_depth_averaged}). This analysis addresses two questions. The first is whether the proposed ensemble system improves deterministic forecast skill relative to the deterministic and persistence baselines. The second is whether it provides a more informative probabilistic forecast than an ensemble constructed by externally imposed spatial perturbations.

Across the five representative variables, including salinity, temperature, zonal current, meridional current, and sea surface height, FuXi-ONS shows consistently strong performance over the full annual forecast range. In deterministic metrics, the ensemble mean of FuXi-ONS generally achieves the lowest RMSE and the highest, or among the highest, ACC over most lead times. The advantage is especially clear relative to persistence. In RMSE, all three data-driven methods substantially outperform persistence across the full forecast range for all five variables, showing that the learned forecasts capture genuine ocean evolution rather than simply retaining the initial state. In ACC, persistence drops rapidly after the early forecast range and remains at a comparatively low level thereafter. FuXi-ONS, by contrast, preserves substantially higher correlation throughout the annual horizon. Compared with the deterministic FuXi-Aim baseline, FuXi-ONS also maintains a systematic advantage across most variables and lead times, indicating that the ensemble framework improves not only uncertainty representation but also the central forecast trajectory itself.

The comparison with FuXi-Aim-Perlin is particularly informative for probabilistic skill. In both CRPS and SSR, only the two ensemble methods are directly comparable, and FuXi-ONS consistently outperforms FuXi-Aim-Perlin across all five variables. In the CRPS curves, FuXi-ONS remains uniformly lower over nearly the entire forecast range, while the gap generally widens with increasing lead time. This behavior shows that the advantage of ensemble forecasting becomes more important as uncertainty accumulates over time, and that the gain of FuXi-ONS cannot be explained by simply adding generic spatial perturbations to a deterministic forecast. Instead, the learned ensemble generation process produces a predictive distribution that is more consistent with the observed forecast uncertainty evolution \cite{Hersbach2000,zhong2024fuxiens,Price2024_GenCast}.

The SSR curves further support this interpretation. FuXi-Aim-Perlin exhibits very low SSR at short lead times and remains substantially below FuXi-ONS throughout the forecast horizon for all variables, indicating that the spread introduced by Perlin perturbations is not commensurate with the actual forecast error. By contrast, FuXi-ONS produces markedly higher SSR and a much more coherent temporal evolution. Although the SSR values of FuXi-ONS still remain below the ideal value, implying that the ensemble is not yet fully calibrated and remains under-dispersive, the improvement over the Perlin-noise baseline is large and consistent. This is the key point here. The objective is not merely to generate visually diverse ensemble members, but to produce uncertainty estimates whose spread varies in a physically meaningful way with forecast difficulty \cite{fortin2014should,zhong2024fuxiens}.

The deterministic metrics also reveal clear differences among variables. In RMSE, FuXi-ONS maintains the best overall performance across all five variables, with the advantage over the other learned baselines becoming more evident at medium and long lead times. In ACC, however, the variable dependence is more distinct. The long-range correlation is better preserved for the zonal current and sea surface height, whereas salinity and temperature show a more gradual but stronger decline toward the later part of the forecast range. The deterministic FuXi-Aim baseline degrades more noticeably at medium-to-long lead times, especially around the 180--240 day range, where its ACC approaches or falls below persistence for several variables. By contrast, the ensemble-based methods retain higher ACC over most of the forecast horizon and remain almost consistently above persistence. This behavior is particularly clear for the current fields and sea surface height.

Overall, Fig.~\ref{fig_depth_averaged} shows that FuXi-ONS improves both deterministic accuracy and probabilistic forecast quality in the depth-averaged sense. In probabilistic evaluation, FuXi-ONS consistently yields a better predictive distribution than FuXi-Aim-Perlin. In deterministic evaluation, its ensemble mean also provides the strongest overall performance among the learned methods, while all data-driven approaches substantially outperform persistence. These results show that FuXi-ONS is not simply a stochastic wrapper around a deterministic ocean forecasting model. Rather, it operates as a genuine data-driven ensemble forecasting system, delivering a more accurate central forecast together with a more realistic characterization of forecast uncertainty over long-range global ocean evolution.

\subsection{Vertical structure of forecast improvement}

\begin{figure}[ht]
\centering
\includegraphics[width=\linewidth]{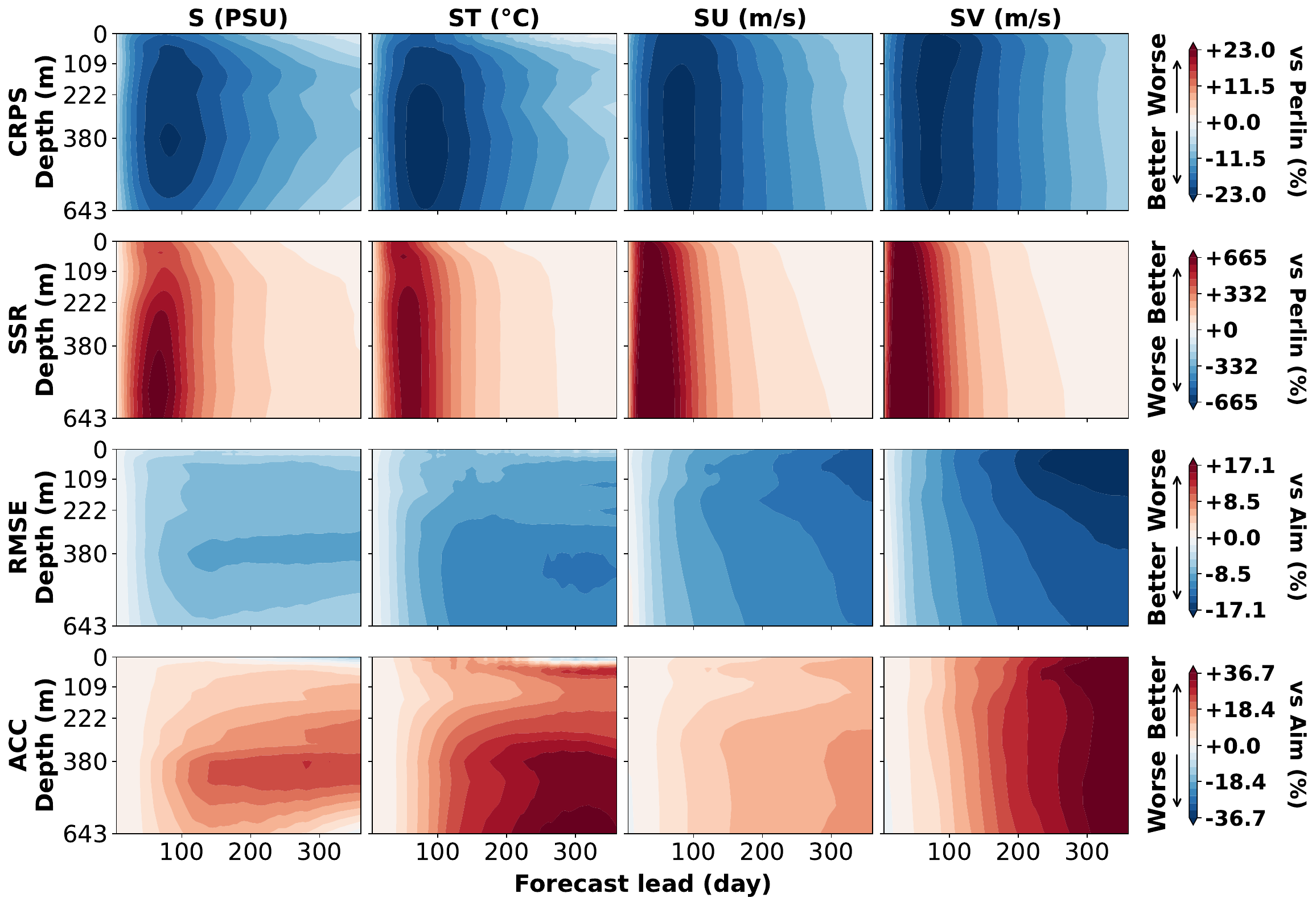}
\caption{\textbf{Depth-dependent normalized improvement of FuXi-ONS as a function of forecast lead time.} Columns correspond to salinity (S), temperature (ST), zonal current (SU), and meridional current (SV). Rows show the relative changes in CRPS, SSR, RMSE, and ACC, respectively, across forecast lead times and depth levels. For CRPS and SSR, the normalized improvement is computed relative to FuXi-Aim-Perlin; for RMSE and ACC, it is computed relative to FuXi-Aim.}
\label{fig_nrmse_vs_fuxi_aim}
\end{figure}
\FloatBarrier
We next examine how the gain of FuXi-ONS is distributed in the vertical dimension relative to the reference baselines (Fig.~\ref{fig_nrmse_vs_fuxi_aim}). Unlike Fig.~\ref{fig_depth_averaged}, which reports depth-averaged skill, this analysis resolves the relative improvement jointly in depth and lead time. The purpose here is not to infer the absolute three-dimensional forecast structure from these fields alone, but to identify where the learned ensemble formulation brings the largest benefits within the water column.

A clear result is that the improvement is vertically organized rather than confined to a narrow surface layer. For probabilistic evaluation, FuXi-ONS yields lower CRPS than FuXi-Aim-Perlin across almost the full depth range for salinity, temperature, and both current components. The reduction is spatially coherent in the vertical and remains visible through a large fraction of the evaluated upper ocean. This point is important because the comparison here is against an already strong and widely used perturbation strategy. Perlin-type multi-scale perturbations can provide an effective generic baseline, but the present results suggest that such externally imposed noise is not sufficient for global three-dimensional ocean forecasting, where the distribution and evolution of uncertainty differ substantially across variables and depths. In this setting, a learned ensemble generation process is better able to capture the relative structure of forecast uncertainty throughout the water column.
The SSR patterns strengthen this interpretation. FuXi-ONS shows systematically higher SSR than FuXi-Aim-Perlin across nearly the entire depth-lead domain, with the most pronounced gains appearing in the middle and deeper layers rather than near the surface. This behavior is especially clear for the current components, where the relative improvement remains strong through much of the subsurface column. Salinity and temperature show the same overall tendency, although with a smoother vertical structure. These results indicate that the advantage of FuXi-ONS over Perlin perturbations becomes particularly important below the immediate surface, where simple multi-scale noise injection is less able to produce spread that remains commensurate with forecast difficulty in a vertically varying ocean state \cite{fortin2014should,zhong2024fuxiens}.

The deterministic comparison against FuXi-Aim shows that the ensemble formulation also improves the central forecast trajectory in a broadly distributed vertical sense. In RMSE, FuXi-ONS reduces error across almost the full depth range for all four three-dimensional variables, with the clearest relative gains appearing in the upper-to-middle layers. The pattern is particularly pronounced for the current fields, but the same conclusion holds for salinity and temperature as well. Thus, the advantage of the ensemble model is not restricted to probabilistic calibration. Relative to the deterministic backbone, it also yields a broadly improved mean prediction across the three-dimensional ocean state.
A similarly consistent picture emerges in ACC. Relative to FuXi-Aim, FuXi-ONS improves correlation through most of the water column for all four three-dimensional variables. The strongest gain appears in the meridional current, while salinity, temperature, and the zonal current also show broadly positive improvements over depth. Importantly, the depth structures seen in ACC are in general consistent with those in RMSE for the same variables. Variables that show broader or stronger relative error reduction also tend to exhibit broader or stronger correlation improvement. This agreement between RMSE and ACC suggests that the gain of FuXi-ONS is not metric-specific. Instead, it reflects a robust improvement in the forecast evolution of the three-dimensional ocean state relative to the deterministic baseline.

Overall, Fig.~\ref{fig_nrmse_vs_fuxi_aim} shows that the advantage of FuXi-ONS is clearly depth dependent. Relative to FuXi-Aim-Perlin, it yields more favorable probabilistic skill across the vertical column. Relative to FuXi-Aim, it also improves deterministic skill for all four three-dimensional variables. The improvement therefore extends beyond depth-averaged scores and appears as a coherent vertical gain relative to both deterministic and noise-perturbed baselines.

\subsection{Comparison with numerical ensemble models}
\begin{figure}[ht]
\centering
\includegraphics[width=\linewidth]{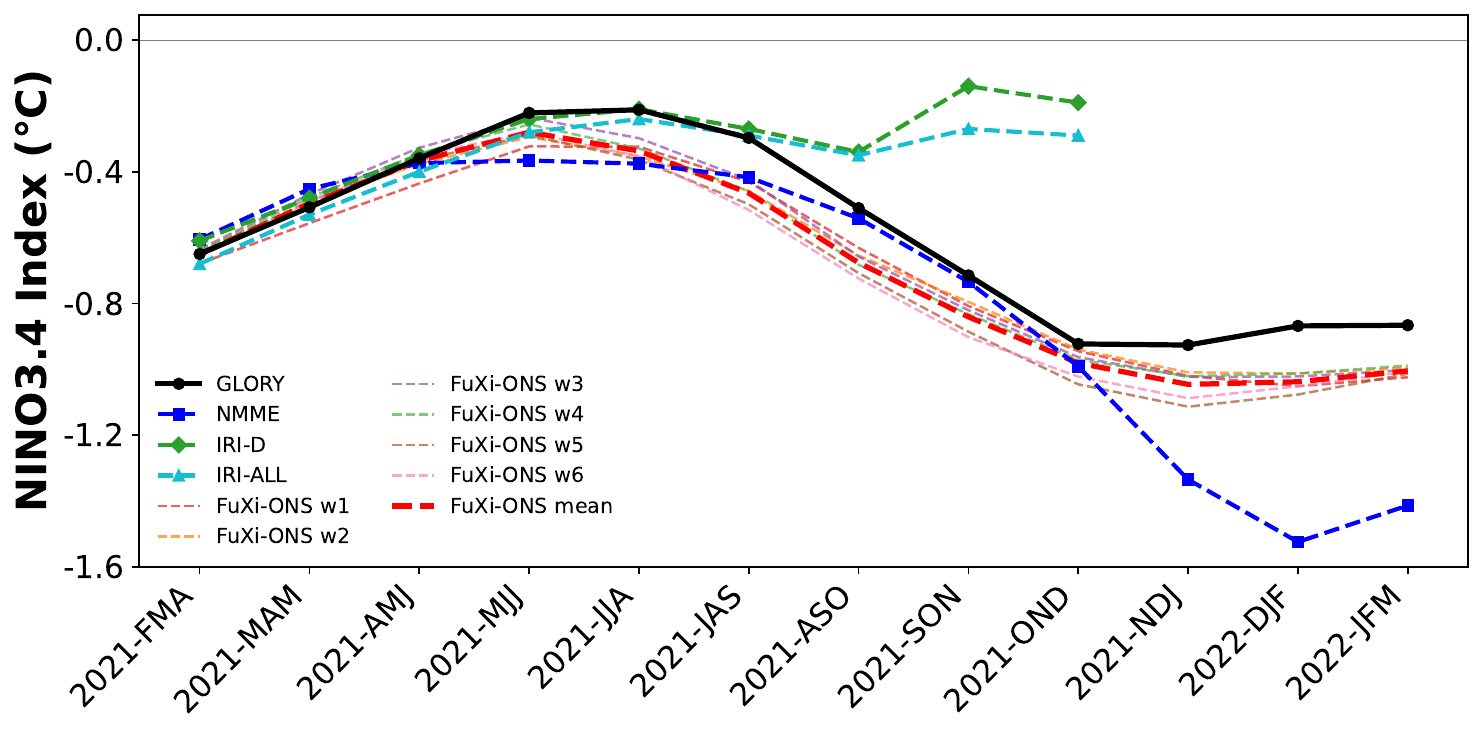}
\caption{\textbf{Ni\~no3.4 forecasts for the 2021-02 initialization.} Ni\~no3.4 index forecasts from GLORY, NMME, IRI-D, IRI-ALL, and FuXi-ONS as a function of lead season. IRI-D uses only dynamical models, whereas IRI-ALL combines dynamical and statistical models. FuXi-ONS w1--w6 denote forecasts initialized every 5 days within the month, specifically 1 February, 6 February, 11 February, 16 February, 21 February, and 26 February 2021, and the FuXi-ONS mean denotes the average over these six forecasts. The horizontal axis is expressed in overlapping 3-month seasons, where each value is the mean over the labeled season, for example FMA denotes the average of February, March, and April.}
\label{fig_nino34_compare_202102}
\end{figure}
\FloatBarrier

\begin{figure}[ht]
\centering
\includegraphics[width=\linewidth]{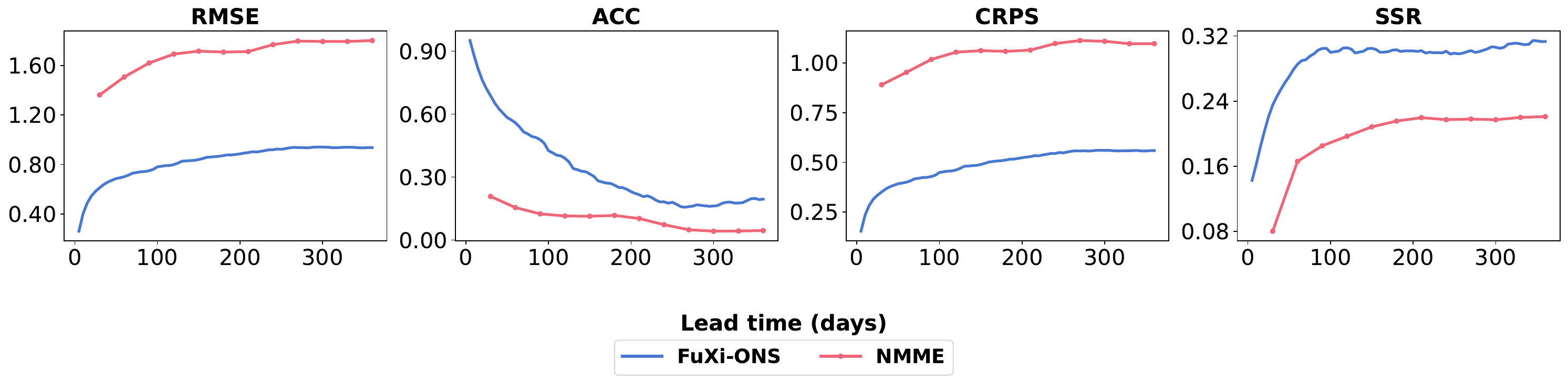}
\caption{\textbf{Comparison of FuXi-ONS with NMME for sea-surface temperature forecasting.} RMSE, ACC, CRPS, and SSR of monthly SST forecasts as a function of lead time. The blue solid line denotes FuXi-ONS, and the red line with circle markers denotes NMME. All metrics are averaged over forecasts initialized in 2021.}
\label{fig_st0_nmme}
\end{figure}
\FloatBarrier
We next compare FuXi-ONS with established seasonal forecast references in the subset of settings where direct comparison is feasible. Because the publicly available reference products used here are monthly and SST-centered, this comparison is restricted to the 2021 initializations and focuses on two related diagnostics: the Ni\~no3.4 index and monthly SST forecast skill. The ENSO comparison includes NMME together with the IRI forecast products, whereas the field-level SST comparison is carried out against NMME, for which the corresponding gridded ensemble output is directly available.

We begin with ENSO, because Ni\~no3.4 provides a compact summary of basin-scale tropical Pacific variability. As shown in Fig.~\ref{fig_nino34_compare_202102}, FuXi-ONS reproduces the seasonal evolution of the 2021 Ni\~no3.4 anomaly more faithfully than the reference seasonal forecast systems. In the first half of the forecast, FuXi-ONS, NMME, IRI-D, and IRI-ALL all capture the gradual weakening of the negative anomaly from FMA to boreal summer. The differences emerge in the second half of the forecast. FuXi-ONS correctly captures the renewed cooling from late summer into winter, while both IRI products remain substantially too warm and fail to reproduce the late-2021 intensification of the cold anomaly. NMME follows the transition more closely at first, but subsequently overshoots the cooling and becomes markedly too cold during NDJ, DJF, and JFM. By contrast, the FuXi-ONS ensemble mean remains much closer to the reference trajectory over these later seasons, although it still shows a moderate cold bias. This comparison is important because ENSO prediction is not merely a local SST problem. It tests whether the forecast system can represent the dominant large-scale mode of seasonal ocean variability in the tropical Pacific.

The monthly SST comparison provides a field-level complement to this index-based result. As shown in Fig.~\ref{fig_st0_nmme}, FuXi-ONS substantially outperforms NMME in RMSE and CRPS across the full forecast range, indicating both lower deterministic error and a markedly better probabilistic forecast distribution. FuXi-ONS also maintains higher ACC than NMME throughout the evaluation period, although the gap narrows at longer lead times. In SSR, both systems remain under-dispersive, but FuXi-ONS consistently achieves larger values, indicating a better match between ensemble spread and forecast error. The SST results therefore support the ENSO comparison above. The advantage of FuXi-ONS is not limited to a scalar climate index, but is also reflected in the monthly SST field from which the Ni\~no3.4 anomaly is derived.

Overall, these comparisons show that FuXi-ONS is competitive with, and in several key diagnostics clearly stronger than, an established numerical seasonal ensemble benchmark in the 2021 SST-centered setting. The Ni\~no3.4 analysis demonstrates that FuXi-ONS better captures the seasonal evolution of ENSO-related variability, while the monthly SST metrics provide consistent field-based support for the same conclusion.

\section{Discussion}
In this study, we introduced FuXi-ONS, to the best of our knowledge the first data-driven ensemble prediction system for global ocean forecasting. FuXi-ONS extends recent progress in AI-based deterministic ocean prediction to the probabilistic setting, where the goal is not only to predict an accurate mean trajectory, but also to represent forecast uncertainty in a physically meaningful way. The results show that this is feasible at global scale. Relative to deterministic and noise-perturbed baselines, FuXi-ONS improves both ensemble-mean skill and probabilistic forecast quality across variables, depths, and lead times. The consistent gains in CRPS and SSR indicate that the learned ensemble formulation does more than increase member diversity. It produces spread that is more closely aligned with flow-dependent forecast difficulty.

These results also suggest that ocean ensemble prediction cannot be reduced to deterministic forecasting plus generic stochastic perturbation. This is most evident in the comparison with FuXi-Aim-Perlin. Perlin-type perturbations provide a simple and effective reference, but their limitations become clear in long-range three-dimensional ocean prediction, where uncertainty is strongly structured across variables and depths. In this setting, externally imposed multi-scale noise is not sufficient to reproduce the vertical and multivariate organization of forecast uncertainty. The advantage of FuXi-ONS suggests that learned, state-dependent perturbations are better suited to this problem. The model benefits further from encoding the initialization atmosphere into a stable forcing representation, which helps preserve large-scale predictability without relying on long autoregressive atmospheric rollouts.

At the same time, FuXi-ONS should not be interpreted as a replacement for numerical ensemble systems. The model is trained on reanalysis products generated by numerical forecasting and data assimilation systems, and therefore builds on the information and physical consistency already contained in those products. Its contribution is different. It shows that, once such a foundation is available, probabilistic global ocean forecasts can be generated much more efficiently than by repeatedly integrating a conventional ensemble model. This makes data-driven ensemble forecasting a practical complement to existing numerical systems rather than an alternative to the full numerical forecasting chain.

Several limitations remain. Although FuXi-ONS substantially improves SSR relative to the Perlin-noise baseline, the ensemble is still under-dispersive in an absolute sense. In addition, the present system operates on a global $1^\circ \times 1^\circ$ grid and 5-day forecast interval, which is appropriate for long-range global prediction but does not resolve finer mesoscale or coastal processes. These limitations point to clear future directions, including improved ensemble calibration, finer spatial and temporal resolution, and tighter integration with data assimilation and hybrid numerical--ML workflows.

Taken together, these results show that global ocean ensemble prediction can be reformulated as a tractable data-driven problem without losing probabilistic meaning. The significance of this work lies not only in improved verification metrics, but in establishing a feasible path toward efficient probabilistic ocean forecasting at global scale.

\begin{figure}[ht]
\centering
\includegraphics[width=\linewidth]{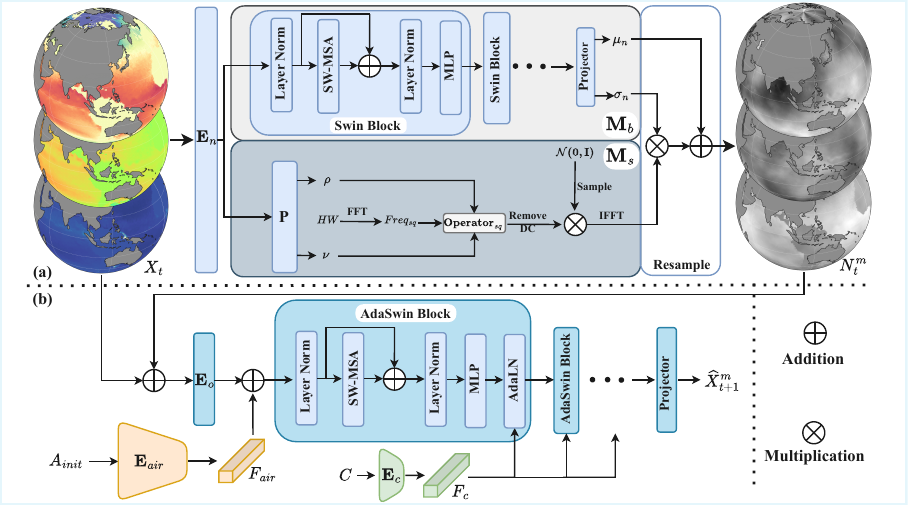}
\caption{\textbf{Architecture of the FuXi-ONS ensemble forecasting framework}. (a) Structured-noise generation module. Given the current ocean state $X_t$, the model predicts state-dependent perturbation statistics and samples spatially correlated noise through an SPDE-based Matérn formulation, yielding the structured perturbation field $N_t^{(m)}$. (b) Forecast module. The initial atmospheric state $A_{\mathrm{init}}$ and auxiliary input $C$ are encoded into latent conditioning features $F_{\mathrm{air}}$ and $F_c$. The perturbed ocean state $X_t + N_t^{(m)}$ is then mapped to the next-step forecast $\hat{X}_{t+1}^{(m)}$. Multiple ensemble members are generated by resampling $N_t^{(m)}$ at each forecast step.}
\label{struc}
\end{figure}
\FloatBarrier
\section{Methods}
\subsection{Data}
GLORYS12 \cite{jean2021copernicus} is a global eddy-rich ocean reanalysis product that provides
dynamically consistent estimates of three-dimensional ocean states. In this
work, GLORYS12 is used to provide both the ocean inputs and the prediction
targets. We use five ocean variables, namely temperature (ST), salinity (S), zonal
current (SU), meridional current (SV), and sea-surface height (SSH). Among them, temperature,
salinity, zonal current, and meridional current are extracted at 23 depth
levels, whereas sea-surface height is defined only at the surface. As a result,
the ocean state contains a total of 93 channels and is written as
$X_t \in \mathbb{R}^{C_o \times H \times W}$ with $C_o = 93$. Here,
$H=180$ and $W=360$, corresponding to a global $1^{\circ}\times1^{\circ}$ grid.
This notation is adopted throughout the paper by flattening the original
four-dimensional organization of ocean data, namely variable, depth, latitude,
and longitude, into a three-dimensional tensor with channel, latitude, and
longitude dimensions.

ERA5 \cite{hersbach2020era5} is the fifth-generation global atmospheric reanalysis produced by the
European Centre for Medium-Range Weather Forecasts. In this work, ERA5 is used
only for atmospheric conditioning. We use eight near-surface atmospheric
variables: 2-m temperature (T2m), 2-m dewpoint temperature (d2m), mean sea
level pressure (msl), surface solar radiation (SSRD), total
precipitation (TP), 10-m zonal wind (u10m), 10-m meridional wind (v10m), and
surface thermal radiation downwards (strd). The atmospheric input is denoted by
$A_{\mathrm{init}} \in \mathbb{R}^{C_a \times H \times W}$ with $C_a = 8$.

The data are divided chronologically into training, validation, and test sets.
The training period spans 1993-01-01 to 2020-06-30, the validation period spans
2020-07-01 to 2020-12-31, and the test period spans 2021-01-01 to 2023-12-31.
Samples are constructed by randomly selecting a day and applying a 5-day moving
average window, rather than averaging over fixed calendar dates. Forecast
targets are defined at 5-day intervals, consistent with the autoregressive
rollout step of the model.
\subsection{Problem formulation and model overview}
FuXi-ONS is a data-driven ensemble forecasting system for multivariate global ocean prediction. Given the ocean state at step $t$, the model predicts the state at step $t+1$ and is autoregressively rolled out at 5-day intervals out to 365 days. Ensemble members are generated by resampling state-dependent perturbations during rollout.

The framework contains three functional components: a structured-noise
generation module, an atmospheric conditioning module, and an ocean forecasting
module. The noise-generation module produces flow-dependent perturbations from
the current ocean state. The atmospheric conditioning module extracts
large-scale forcing information from the initialization atmosphere and provides
stable atmospheric context throughout the forecast. The forecasting module then
combines the perturbed ocean state with atmospheric and auxiliary conditioning
features to predict the next ocean state.

Let $X_t \in \mathbb{R}^{C_o \times H \times W}$ denote the ocean state at step
$t$, $A_{\mathrm{init}} \in \mathbb{R}^{C_a \times H \times W}$ denote the
atmospheric input at the initialization time, and
$C \in \mathbb{R}^{C_c \times H \times W}$ denote the auxiliary conditioning
inputs defined on the physical grid. To improve computational efficiency,
FuXi-ONS maps full-resolution inputs on the physical grid of size
$H \times W$ to latent feature maps of size $h \times w$, where $h < H$ and
$w < W$. Specifically, the modules $\mathrm{E}_n$ and $\mathrm{E}_o$ are
convolutional embedding layers that project the perturbation-related input and
the ocean input, respectively, onto this latent feature grid. The atmospheric
conditioning feature $F_{\mathrm{air}}$ and the auxiliary conditioning feature
$F_c$ are also represented on the latent grid, with
$F_{\mathrm{air}} \in \mathbb{R}^{C_f \times h \times w}$ and
$F_c \in \mathbb{R}^{C_c' \times h \times w}$.

For ensemble member $m$, FuXi-ONS predicts the next-step ocean state as
\begin{equation}
\hat{X}_{t+1}^{(m)} =
\mathrm{M}_f\bigl(X_t + N_t^{(m)},\, F_{\mathrm{air}},\, F_c \bigr),
\label{eq:1}
\end{equation}
where $N_t^{(m)} \in \mathbb{R}^{C_o \times H \times W}$ is the structured
perturbation field for member $m$ at step $t$, and $\mathrm{M}_f(\cdot)$
denotes the ocean forecasting network. Multiple ensemble members are obtained
by resampling $N_t^{(m)}$ at each forecast step, and the final latent prediction
is mapped back to the physical grid through a deconvolution-based projector.
\subsection{Structured noise generation}

A central challenge in data-driven ocean ensemble forecasting is to generate
ensemble perturbations that are both diverse and physically plausible. Ocean
uncertainty is strongly structured in space, with correlated variability across
multiple scales. As a result, independent pixel-wise perturbations are generally
inconsistent with the organization of ocean states and tend to produce
unrealistic ensemble members.

To address this issue, FuXi-ONS generates structured perturbations by combining state-dependent amplitude prediction with an SPDE-based spectral sampler. The current ocean state $X_t$ is first embedded by $\mathrm{E}_n$ and processed by the noise backbone $\mathrm{M}_b(\cdot)$ to predict the perturbation statistics $\mu_n, \sigma_n \in \mathbb{R}^{C_o \times H \times W}$,
which determine the location and scale of the perturbation field. In parallel, a lightweight parameter network $\mathrm{P}(\cdot)$, implemented as a stack of convolutional layers, predicts two channel-wise parameters $\rho_t, \nu_t \in \mathbb{R}^{C_o},$ where each channel is assigned one correlation-range parameter and one smoothness parameter. In the implementation, these parameters are broadcast from shape $C_o$ to $C_o \times 1 \times 1$ and then applied to all spatial locations within the corresponding channel.

Inspired by the SPDE formulation of Mat\'ern-type random fields
\cite{lindgren2011spde}, FuXi-ONS samples structured noise in the spectral domain. For channel $c$, we define $\kappa_{t,c} = \rho_{t,c}^{-1}$, where $\rho_{t,c}$ controls the correlation range, and $\nu_{t,c}$ controls the spectral smoothness. Let $k$ denote the spatial wavenumber and
$\hat{W}_c(k)$ denote complex Gaussian white noise in the Fourier domain. The
channel-wise structured noise is generated as
\begin{equation}
\hat{u}_{t,c}(k)
=
\hat{W}_c(k)
\left(
\kappa_{t,c}^2 + 4\pi^2 \lVert k \rVert^2
\right)^{-\nu_{t,c}/2},
\label{eq:spectral_spde}
\end{equation}
followed by inverse Fourier transform to obtain
$u_t^{(m)} \in \mathbb{R}^{C_o \times H \times W}$. In practice, the zero
frequency component is removed to avoid introducing a global mean shift, and
the sampled field is normalized independently for each channel.

The final perturbation applied to ensemble member $m$ at step $t$ is
\begin{equation}
N_t^{(m)} = \mu_n + \sigma_n \odot u_t^{(m)},
\label{eq:noise}
\end{equation}
where $\odot$ denotes element-wise multiplication. In this way, the amplitude
of the perturbation is predicted as a full spatial field, while its correlation
range and smoothness are controlled by channel-wise parameters estimated from
the current ocean state. This design allows FuXi-ONS to generate ensemble
spread that is spatially coherent, flow dependent, and variable specific.

Noise is resampled at every forecast step, so ensemble diversity evolves
throughout the autoregressive rollout rather than being determined only at
initialization.
\subsection{Atmospheric conditioning}
Ocean prediction on seasonal to annual timescales remains strongly influenced by atmospheric surface forcing. Directly coupling the model to rolling atmospheric forecasts is problematic, however, because atmospheric forecast errors accumulate rapidly at extended lead times.

FuXi-ONS therefore conditions the ocean forecast on the initial atmospheric
state rather than on autoregressively updated atmospheric inputs. Specifically,
the atmospheric encoder $\mathrm{E}_{\mathrm{air}}(\cdot)$ maps
$A_{\mathrm{init}} \in \mathbb{R}^{C_a \times H \times W}$ to a latent
conditioning feature
\begin{equation}
F_{\mathrm{air}} = \mathrm{E}_{\mathrm{air}}(A_{\mathrm{init}}),
\label{eq:fair}
\end{equation}
where $F_{\mathrm{air}} \in \mathbb{R}^{C_f \times h \times w}$. The encoder
consists of a convolutional embedding layer followed by stacked Swin blocks,
which compress the full-resolution atmospheric input into a compact latent
representation while preserving large-scale forcing information relevant to
ocean evolution.

To regularize the atmospheric conditioning branch, we introduce an atmospheric
consistency loss. Let $A_t \in \mathbb{R}^{C_a \times H \times W}$ denote the
ERA5 atmospheric state at the valid time corresponding to forecast step $t$.
Its encoded representation is
\begin{equation}
F_{\mathrm{air},t}^{\ast} = \mathrm{E}_{\mathrm{air}}(A_t),
\label{eq:fair_true}
\end{equation}
where $F_{\mathrm{air},t}^{\ast} \in \mathbb{R}^{C_f \times h \times w}$. We
then require the conditioning feature extracted from the initialization
atmosphere to remain close to the feature extracted from the true atmosphere at
the corresponding valid time. The atmospheric consistency loss is defined as
\begin{equation}
L_{\mathrm{air}} =
\frac{1}{C_f h w}
\left\|
F_{\mathrm{air}} - F_{\mathrm{air},t}^{\ast}
\right\|_1.
\label{eq:lair}
\end{equation}
This term acts as a regularizer that encourages the atmospheric conditioning
features to retain temporally meaningful large-scale forcing information.
\subsection{Forecasting backbone and Auxiliary conditioning}

In addition to atmospheric forcing, FuXi-ONS uses auxiliary conditioning inputs
to encode static and temporal information relevant to ocean prediction. Let
$C \in \mathbb{R}^{C_c \times H \times W}$ denote the auxiliary conditioning
inputs on the physical grid, including longitude--latitude coordinates,
land--sea mask, forecast lead time, time of day, and day of year. These inputs
are mapped to a latent feature by the auxiliary encoder
\begin{equation}
F_c = \mathrm{E}_c(C),
\label{eq:fc}
\end{equation}
where $F_c \in \mathbb{R}^{C_c' \times h \times w}$. The encoder
$\mathrm{E}_c(\cdot)$ maps the auxiliary inputs from the physical grid to the
latent feature grid used by the forecasting network.

The ocean forecasting network $\mathrm{M}_f(\cdot)$ takes as input the
perturbed ocean state $X_t + N_t^{(m)}$, the atmospheric conditioning feature
$F_{\mathrm{air}}$, and the auxiliary conditioning feature $F_c$. The ocean
input is first embedded by $\mathrm{E}_o(\cdot)$ into the latent feature space,
after which the latent ocean feature is combined with $F_{\mathrm{air}}$ and
processed by the forecasting backbone. The backbone consists of 20 stacked
AdaSwin blocks. Each AdaSwin block is built upon a standard Swin block, while
the auxiliary conditioning feature $F_c$ is injected through adaptive layer
normalization (AdaLN). In this way, $F_c$ provides conditioning signals to
modulate the intermediate ocean features throughout the forecasting network.

After passing through the stacked AdaSwin blocks, the latent prediction is
mapped back to the physical grid by a deconvolution-based projector, yielding
the next-step ocean prediction
$\hat{X}_{t+1}^{(m)} \in \mathbb{R}^{C_o \times H \times W}$.

\subsection{Prediction loss and training objective}
The main supervision term is a latitude-weighted Charbonnier loss \cite{Charbonnier1994} computed over
ocean grid cells. Let $\hat{X}_{t+1}^{(m)} \in \mathbb{R}^{C_o \times H \times W}$
denote the one-step prediction for ensemble member $m$, and let
$X_{t+1} \in \mathbb{R}^{C_o \times H \times W}$ denote the corresponding target
ocean state. The one-step prediction loss is defined as
\begin{equation}
L_{\mathrm{pred}}^{(1)} =
\frac{1}{C_o H W}
\sum_{c=1}^{C_o}
\sum_{i=1}^{H}
\sum_{j=1}^{W}
\omega_i
\sqrt{
\left(
\hat{X}_{t+1,c,i,j}^{(m)} - X_{t+1,c,i,j}
\right)^2 + \epsilon^2
},
\label{eq:lpred1}
\end{equation}
where $\omega_i$ is the latitude weight at latitude index $i$, proportional to
the cosine of latitude, and $\epsilon$ is a small constant for numerical
stability. This formulation accounts for the unequal area represented by
latitude--longitude grid cells and provides a robust regression loss for
multivariate ocean prediction.

During training, FuXi-ONS is optimized autoregressively by unrolling the forecast
for $K=10$ steps, corresponding to 50 days. Let
$\hat{X}_{t+k}^{(m)} \in \mathbb{R}^{C_o \times H \times W}$ denote the prediction
at autoregressive step $k$. The rollout prediction loss is defined as
\begin{equation}
L_{\mathrm{pred}} =
\frac{1}{K}
\sum_{k=1}^{K}
L_{\mathrm{pred}}^{(k)},
\label{eq:lpred}
\end{equation}
where each $L_{\mathrm{pred}}^{(k)}$ is computed using the same
latitude-weighted Charbonnier form as Eq.~(\ref{eq:lpred1}) at forecast step
$t+k$.

The total training objective is
\begin{equation}
L_{\mathrm{total}} = L_{\mathrm{pred}} + \lambda_{\mathrm{air}} L_{\mathrm{air}},
\label{eq:ltotal}
\end{equation}
where $\lambda_{\mathrm{air}}$ controls the strength of the atmospheric
consistency regularization.

FuXi-ONS is trained end to end using AdamW with cosine-annealed learning-rate
decay and linear warm-up. Distributed training is performed with Fully Sharded
Data Parallel and BF16 mixed precision. To improve long-range forecast stability,
we adopt progressive curriculum training, in which the autoregressive rollout
horizon is gradually increased during training until reaching 10 forecast steps.

\subsection{Evaluation metrics}

Deterministic forecast skill is evaluated using root-mean-square error (RMSE)
and anomaly correlation coefficient (ACC). Probabilistic forecast quality is
assessed using ensemble spread, continuous ranked probability score (CRPS), and
rank histograms. We further examine spread--skill consistency by comparing the
evolution of ensemble spread and forecast error across lead times and regions.
For climate-scale evaluation, we compute the Ni\~no 3.4 index from the predicted
sea-surface temperature fields to assess ENSO forecast skill.

\backmatter

\bmhead{Supplementary information}

\begin{itemize}
  \item[] Supplementary Text S1 to S10
  \item[] Figs S1 to S19
  \item[] References
\end{itemize}

\bmhead{Acknowledgements}




\bibliography{sn-bibliography}

\end{document}